\documentclass[aip,jcp,showpacs,preprintnumbers,amsmath,amssymb,twocolumn,reprint]{revtex4-1}
\usepackage{amsfonts}

\usepackage{graphicx}
\usepackage{dcolumn}
\usepackage{bm}
\usepackage{amssymb}
\usepackage{array}
\usepackage{longtable}

\usepackage{hhline}

\begin{document}

\preprint{AIP/123-QED}

\title{Laplacian spectra of recursive treelike small-world polymer networks:\\ Analytical solutions and applications}

\author{Hongxiao Liu}

\author{Zhongzhi Zhang}
\email{zhangzz@fudan.edu.cn}
\homepage{http://www.researcherid.com/rid/G-5522-2011}

\affiliation {School of Computer Science, Fudan University, Shanghai
200433, China}

\affiliation {Shanghai Key Lab of Intelligent Information
Processing, Fudan University, Shanghai 200433, China}




\date{\today}

\begin{abstract}
A central issue in the study of polymer physics is to understand the relation between the geometrical  properties of macromolecules and various dynamics, most of which are encoded in the Laplacian spectra of a related graph describing the macrostructural structure. In this paper, we introduce a family of treelike polymer networks with a parameter, which has the same size as the Vicsek fractals modeling regular hyperbranched polymers. We study some relevant properties of the networks and show that they have an exponentially decaying degree distribution and exhibit the small-world behavior. We then study the Laplacian eigenvalues and their corresponding eigenvectors of the networks under consideration, with both quantities being determined through the recursive relations deduced from the network structure. Using the obtained recursive relations we can find all the
eigenvalues and eigenvectors for the networks with any size. Finally, as some applications, we use the eigenvalues to study analytically or semi-analytically three dynamical processes occurring in the networks, including random walks, relaxation dynamics in the framework of generalized Gaussian structure, as well as the fluorescence depolarization under quasiresonant energy transfer. Moreover, we compare the results with those corresponding to Vicsek fractals, and show that the dynamics differ greatly for the two network families, which thus enables us to distinguish between them.
\end{abstract}



\maketitle

\section{introduction}

A fundamental issue in the study of complex systems is to unveil how the structural properties affect various dynamics, many of which are related to the exact knowledge of the eigenvalues and eigenvectors of Laplacian matrix. Examples include relaxation dynamic in the framework of generalized Gaussian structure (GGS)~\cite{GuBl05}, fluorescence depolarization by quasiresonant energy transfer~\cite{BlVoJuKo05JOL,BlVoJuKo05}, standard discrete-time random walks~\cite{WuZhCh11}, and continuous-time quantum walks~\cite{MuVoBl05,LiZhXuWu11}, and so on. In addition to dynamical processes, Laplacian eigenvalues and eigenvectors are also relevant to diverse structural aspects of complex systems, such as spanning trees~\cite{TzWu00} and resistance distance~\cite{Wu04}. Thus, it of theoretical interest and practical importance to derive exact analytical expressions of Laplacian eigenvalues and eigenvectors for complex systems, which can lead to extensive insights in the contexts of topologies and dynamics.

Given the wide range of applicability, the study of Laplacian  eigenvalues and eigenvectors has been subject of considerable research endeavor for the past few decades. Thus far, the Laplacian eigenvalues for some classes of graphs have been determined exactly, including regular hypercubic lattices~\cite{GuBl05,DejoMe98}, dual Sierpinski gaskets~\cite{CoKa92,MaMaPe97}, Vicsek fractals~\cite{JaWuCo92,JaWu94}, dendrimer also known as Cayley tree~\cite{CaCh97}, and Husimi cacti~\cite{GaBl07,Ga10}. Recent empirical research indicated that some real-life networks (e.g., power grid) display small-world behavior~\cite{WaSt98,AmScBaSt00}. Moreover, these networks are simultaneously characterized by an exponentially decaying degree distribution~\cite{AmScBaSt00}, which cannot be described by above-mentioned networks. However, related work about Laplacian eigenvalues and eigenvectors for small-world exponential networks is much less, notwithstanding the ubiquitous nature of such systems.

In this paper, we define a category of treelike polymer networks controlled by a parameter, which is built in an iterative way. The networks have the same size as that of Vicsek fractals~\cite{Vi83,ZhZhChYiGu08} corresponding to the same parameter and iteration. According to the construction, we study some structural properties of the networks, showing that they have an exponentially decaying degree distribution, and display the small-world property. Moreover, the networks can be assortative, uncorrelated, or disassortative, relying on the parameter. Then, by applying the technique of graph theory and an algebraic iterative procedure, we study the Laplacian eigenvalues and eigenvectors of the networks, obtaining recursive relations for the eigenvalues and eigenvectors, which allow for determining exactly the full eigenvalues and eigenvectors of networks of arbitrary size.

In the second part of this work, by making use of the obtained Laplacian eigenvalues, we study three classic dynamics for the small-world polymer networks, such as trapping with a single trap, relaxation dynamics in the GGS framework, and the fluorescence depolarization under quasiresonant energy transfer. For the trapping problem, we study two particular cases: in the first case the trap is fixed at the central node, while in the other case the trap is distributed uniformly. For both cases, we derive explicit formulas for the average trapping time
and obtain their leading scalings, which follow different behaviors, showing that the position of trap has a substantial effect on the trapping efficiency. For the GGS, we determine three interesting quantities related to the relaxation dynamics, i.e., the averaged monomer displacement, storage module and loss module. Finally, we display the behavior of the fluorescence depolarization. For the three dynamics, we also present a comparison for the behaviors between the small-world polymer networks and Vicsek fractals, and show that they differ strongly.

\section{Network construction and properties}\label{Section2}

In this section, we first introduce a family of treelike small-world polymer networks with an exponential degree distribution, then we study some relevant properties of the networks.

\subsection{Construction method}\label{Section21}

\begin{figure}[h]
\centering
\includegraphics[width=0.8\linewidth,trim=0 0 0 0]{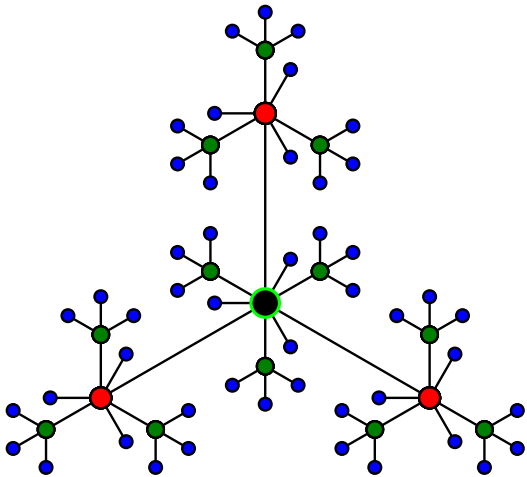}
\caption{(Color online) Construction of a special network corresponding to $f=3$.} \label{graph1}
\end{figure}

The networks being studied have a treelike structure, and are constructed in a deterministically iterative way. Let $U_{g}$ ($g \geq 0$) denote the networks after $g$ iterations. For $g=0$, $U_{0}$ consists of an isolated node, called the central node. For $g=1$, $f$ ($f$ is a positive integer) new nodes are generated connecting the central node to form $U_{1}$. For $g \ge 1$, $U_g$ is obtained from $U_{g - 1}$ by attaching $f$ new nodes to each node in $U_{g-1}$.
Figure~\ref{graph1} illustrates schematically the first several iterative construction processes of a particular network for the case of $f=3$.

According to the construction approach, it is easy to derive that at each iterative step $g_i$ ($g_i \ge 1$), the number of newly generated nodes is $L({g_i}) = f{(f
+ 1)^{{g_i} - 1}}$. Then the total number of nodes at each generation $g$ is
\begin{equation}\label{1f1}
N_g = 1+\sum_{g_i = 1}^g L(g_i)=(f + 1)^g\,,
\end{equation}
and  the total number of edges in $U_{g }$ is ${E_g} = {N_g} - 1 = {(f + 1)^g} - 1$.

\begin{figure}[h]
\centering
\includegraphics[width=0.8\linewidth,trim=0 0 0 0]{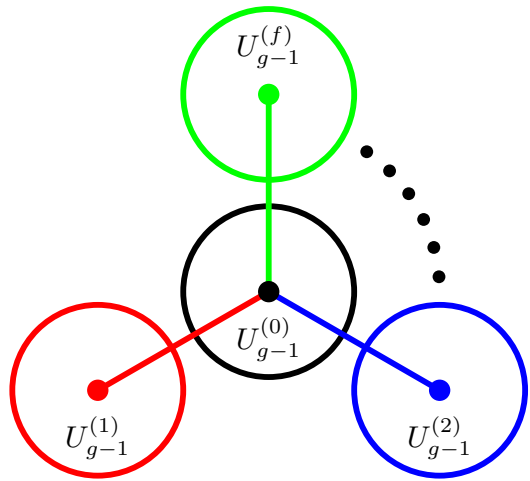}
\caption{(Color online) Second construction method of the small-world polymer networks. $U_g$ consists of $f+1$ copies of $U_{g-1}$, denoted by $U_{g - 1}^{(0)}$, $U_{g - 1}^{(1)}$, $U_{g-1}^{(2)}$,$\ldots$,$U_{g - 1}^{(f)}$, which are connected to each other to form $U_g$ by adding a new edge between the central node of each $U_{g - 1}^{(i)}$ ($i=1,2,\ldots,f$) and the central node  of $U_{g - 1}^{(0)}$. }\label{self-simi}
\end{figure}

In fact, the networks being studied are self-similar, which can be seen from another construction approach. As will be shown below, the central node of $U_g$ has the largest degree, we thus also call it hub node.  Let $h_g$ denote the central node of $U_g$. Then, $U_g$ can be constructed alternatively as follows, highlighting its self-similarity, see Fig.~\ref{self-simi}. To generate $U_g$, we create $f+1$ replicas of $U_{g-1}$, and label them as $U_{g - 1}^{(0)}$, $U_{g - 1}^{(1)}$, $U_{g-1}^{(2)}$,$\ldots$, $U_{g - 1}^{(f)}$, respectively. Moreover, let $h_{g - 1}^{(x)}$ ($x=0,1,2,\ldots,f$) denote the hub of the $U_{g - 1}^{(x)}$. Then, for each $U_{g - 1}^{(i)}$ ($i=1,2,\ldots,f$), we introduce an additional edge connecting its hub node $h_{g - 1}^{(i)}$ to the node $h_{g - 1}^{(0)}$. Thus, through the two steps of replication and connection, we obtain $U_g$ with $h_{g - 1}^{(0)}$ being its hub.

Note that the numbers of nodes and edges of the networks under consideration are identical to those corresponding to Vicsek fractals~\cite{Vi83,ZhZhChYiGu08}, but their structural properties differ greatly from those of Vicsek fractals, as we will show.

\subsection{Structural properties}

We proceed to present some important structural properties
of $U_{g}$, including degree distribution, average path length, diameter,
and degree correlations.

\subsubsection{Degree distribution}

For a network, its degree distribution $P(k)$ is defined as the
probability that a randomly chosen node has a degree of $k$. Let $k_i(g)$ be the degree of node $i$ in $U_{g}$. Assume that node $i$ entered the networks at generation $g_i$ (${g_i} > 0$), then ${k_i}(g_i) = 1$. By construction, at each subsequent iteration, $f$ new nodes will be generated linking to node $i$. Thus, the degree of node $i$ evolves as
\begin{equation}\label{2f1}
{k_i}(g) = {k_i}(g - 1) + f\,.
\end{equation}
Considering  ${k_i}(g_i) = 1$, Eq.~(\ref{2f1}) is solved to yield
\begin{equation}\label{2f2}
{k_i}(g) = 1 + f(g - {g_i})\,,
\end{equation}
which provides the degrees of all nodes except the central one. We label the initial central node by 0; then the degree of node 0 in $U_{g}$ is
\begin{equation}\label{2f3}
{k_0}(g) = fg\,,
\end{equation}
which is the highest among all nodes.

Equations~(\ref{2f2}) and~(\ref{2f3}) show that the degree spectrum of $U_{g}$ is discrete and that all nodes generated at the same generation have the same
degree. Thus, in $U_{g}$, the number of possible node degrees is $g+1$, which is in sharp contrast to that for Vicske fractals, where only three types of degrees exist, that is, 1, 2 and $f$. It follows that the cumulative degree distribution~\cite{Ne03} of the networks addressed is given by
\begin{equation}\label{2f4}
P_{\rm cum}(k) = \sum\limits_{k' = k}^\infty  {P(k')}
\end{equation}
Using Eq.~(\ref{2f2}), we have $P_{\rm cum}(k)=\sum_{k'=k}^{\infty}P(k')=
P\left (g'\leq \phi=g-\frac{k - 1}{f}\right)$. Hence,
\begin{equation}\label{2f41}
P_{\rm cum}(k) = \sum\limits_{g' = 0}^\phi  {\frac{{L(g')}}{{{N_g}}}}  = \frac{{{{(f + 1)}^{g - \frac{{k - 1}}{f}}}}}{{{{(f + 1)}^g}}}
= (f + 1)^{-\frac{k - 1}{f}}\,,
\end{equation}
which decays exponentially with $k$. It is the same with degree distribution $P(k)$, see~\cite{Ne03} for explanation.

\subsubsection{Average path length}

The average path length represents the average of length of the shortest path between two nodes over all node pairs. Assume that each edge in $U_g$ has a unit length. Then the length of the shortest path between nodes $i$ and $j$ in $U_g$, denoted by $d_{ij}(g)$, is the minimum length for the path connecting the two nodes. Let $\bar{d}_g$ represent the average path length of $U_g$, defined by:
\begin{equation}\label{APL01}
\bar{d}_g =\frac{S_{\rm tot}(g)}{N_g(N_g-1)/2}\,,
\end{equation}
where $S_{\rm tot}(g)$ is the sum of $d_{ij}(g)$
over all pairs of nodes, i.e.,
\begin{equation}\label{APL02}
S_{\rm tot}(g) = \sum_{i\neq j} d_{ij}(g)\,.
\end{equation}
We note that in Eq.~(\ref{APL02}), for a pair of nodes $i$ and
$j$ ($i\neq j$), we only count $d_{ij}(g)$ or $d_{ji}(g)$, not both.

Let $\bar{\Theta}_{g}$  and $\Theta_{g}$  the
sets of nodes generated at iteration $g$ or earlier, respectively. Then $S_{\rm tot}(g)$ can be recast as
\begin{small}
\begin{equation}\label{APL03}
S_{\rm tot}(g) = \sum_{i \in \bar{\Theta}_{g},\,j\in \Theta_{g}} d_{ij}(g)+\sum_{i \in \bar{\Theta}_{g},\,j\in \bar{\Theta}_{g}} d_{ij}(g)+\sum_{i \in \Theta_{g},\,j\in \Theta_{g}} d_{ij}(g),
\end{equation}
\end{small}
It is evident that the third term on the right-hand side (rhs) of Eq.~(\ref{APL03}) is exactly $S_{\rm tot}(g-1)$, i.e.,
\begin{equation}\label{APL04}
\sum_{i \in \Theta_{g},\,j\in \Theta_{g}} d_{ij}(g)=S_{\rm tot}(g-1)\,.
\end{equation}
For the first two terms on the rhs of Eq.~(\ref{APL03}), according to the first network construction method, they can be evaluated as
\begin{equation}\label{APL05}
\sum_{i \in \bar{\Theta}_{g},\,j\in \Theta_{g}} d_{ij}(g)=f\,\left[(N_{g-1})^{2}+2\,S_{\rm tot}(g-1)\right],
\end{equation}
and
\begin{equation}\label{APL06}
\sum_{i \in \bar{\Theta}_{g},\,j\in \bar{\Theta}_{g}} d_{ij}(g)=f^{2}\,S_{\rm tot}(g-1)+f\,N_{g-1}(f\,N_{g-1}-1)\,,
\end{equation}
respectively.

Plugging Eqs.~(\ref{APL04}-\ref{APL06}) into Eq.~(\ref{APL03}) leads to
\begin{small}
\begin{eqnarray}\label{APL07}
S_{\rm tot}(g)&=& (f+1)^{2}\,S_{\rm tot}(g-1)+f\,(f+1)(N_{g-1})^{2}-f\,N_{g-1}\nonumber \\
&=& (f+1)^{2\,g}S_{\rm tot}(0) + f\,(f+1)\,\sum_{i=0}^{g-1} \left[(f+1)^{2\,(g-1-i)} (N_i)^{2}\right] \nonumber \\ &\quad&-f\,\sum_{i=0}^{g-1} \left[(f+1)^{2\,(g-1-i)} N_i \right]
\end{eqnarray}
\end{small}
Substituting $S_{\rm tot}(0) = 0$ and $N_i = (f+1)^{i}$ into Eq.~(\ref{APL07}), we can obtain the exact expression for $S_{\rm tot}(g)$ as
\begin{equation}\label{APL08}
S_{\rm tot}(g) =(fg-1){(f+1)^{2g - 1}} + {(f+1)^{g - 1}}\,.
\end{equation}

Inserting Eq.~(\ref{APL08}) into Eq.~(\ref{APL01}) gives
\begin{eqnarray}\label{APL09}
\bar{d}_g &=& \frac{{(fg - 1){{(f+1)}^{2g - 1}} + {{(f+1)}^{g - 1}}}}{{{{(f+1)}^g}[{{(f+1)}^g} - 1]/2}}\nonumber\\
 &=& \frac{{2(fg - 1){{(f+1)}^g} + 2}}{{{{(f+1)}^{g + 1}} - (f+1)}}\,.
\end{eqnarray}
Recalling $N_g=(f+1)^g$ as given in Eq.~(\ref{1f1}), we
have $g=\ln N_g / \ln (f+1)$, both of which enable us to write $\bar{d}_g$ in term of network size $N_g$ as
\begin{eqnarray}\label{APL10}
\bar{d}_g &=& \frac{{2(f\,\ln {N_g}/\ln (f+1) - 1){N_g} + 2}}{{(f+1){N_g} - (f+1)}}\nonumber\\
 &=& \frac{{2f}}{{(f+1)\ln (f+1)}}\frac{{{N_g}\ln {N_g}}}{{{N_g} - 1}} + \frac{2}{{f+1}}\frac{1}{{{N_g} - 1}}\,.
\end{eqnarray}
When the network size is large enough, we have
\begin{equation}\label{APL11}
\bar{d}_g \cong \frac{{2f}}{{(1 + f)\ln (1 + f)}}\ln {N_g}\,,
\end{equation}
which increases logarithmically with the network size $g$, showing that the networks display the small-world behavior~\cite{WaSt98}.

\subsubsection{Diameter}

We have shown that the treelike polymer networks are small-world, since their
average path length grows as a logarithmic function of network size. In addition to average path length, sometimes, diameter is also used to characterize the small-world phenomenon, since small diameter is consistent with the concept of small-world. For a network, its diameter is defined as the maximum of the shortest distances between all pairs of nodes in the network. Let ${\rm diam} (U_g)$ denote the diameter of $U_g$, below we will compute analytically ${\rm diam} (U_g)$ and show that it also scales logarithmically with the network size.

Clearly, at step $g=1$, ${\rm diam} (U_1)$ equals 2. At each iteration
$g\geq 1$, we call newly-generated nodes at this iteration \emph{active
nodes}. Since all active nodes are connected to those nodes existing
in $U_{g-1}$, it is easy to see that the maximum distance between
an arbitrary active node and those nodes in $U_{g-1}$ is not more than
${\rm diam} (U_1)+1$ and that the maximum distance between any pair of
active nodes is at most ${\rm diam} (U_1)+2$. Hence, at any iteration, the
diameter of the network increases by 2 at most. Then we get $2(g+1)$
as the diameter of $U_g$, which is equal to $2(\log_{f+1}N_g+1)$ growing
logarithmically with the network size. This again indicates that the networks under study are small-world.

\subsubsection{Degree correlations}

For a network, its degree correlations~\cite{Ne02} can be described by the Pearson correlation coefficient $r$, which is in the interval $[-1,1]$. If the network is uncorrelated, $r$ equals zero. Disassortative networks have $r < 0$, while
assortative graphs have $r > 0$. Let $r(f,g)$ be the Pearson degree correlation coefficient of $U_g$. By definition, $r(f,g)$ is given by
\begin{equation}\label{2f7}
r(f,g)= \frac{{ E_g \sum\limits_i {{j_i}{k_i}}  - {{\left[ {\sum\limits_i {\frac{1}{2}({j_i} + {k_i})} } \right]}^2}}}{{ E_g \sum\limits_i {\frac{1}{2}(j_i^2 + k_i^2) - {{\left[ {\sum\limits_i {\frac{1}{2}({j_i} + {k_i})} } \right]}^2}} }}\,,
\end{equation}
where where $j_i$ and $k_i$ are the degrees of the nodes at the two ends of the $i$th edge in $U_g$, where $i \in \{1, 2,\ldots, E_g\}$.

\begin{figure}[h]
\centering
\includegraphics[width=1.0\linewidth,trim=10 0 0 0]{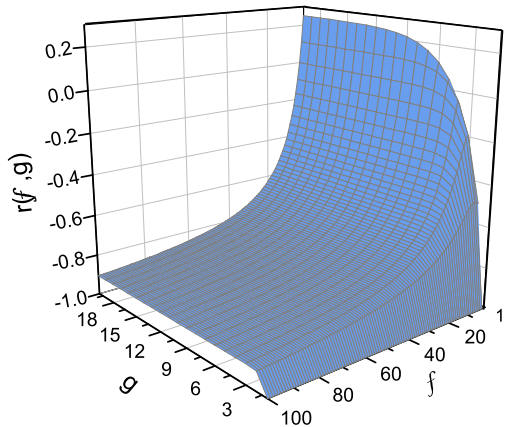}
\caption{(Color online) Pearson correlation coefficient $r(f,g)$ of $U_g$ as a function of $f$ and $g$.}\label{PeaCoe}
\end{figure}

The three terms in numerator and denominator in Eq.~(\ref{2f7}) can be evaluated as
\begin{equation}\label{2f8}
\sum_i j_i\,k_i =(3f+7) (f+1)^g- 2f^2 g^2- 7fg - 3f - 7\,,
\end{equation}
\begin{equation}\label{2f9}
\sum_i \frac{1}{2}(j_i + k_i)= -\frac{1}{2}(f+5)(f+1)^g+ 2fg+\frac{1}{2}(f+5)\,,
\end{equation}
and
\begin{eqnarray}\label{2f10}
\sum_i {\frac{1}{2}(j_i^2 + k_i^2)}&=&\frac{1}{2}(f^2+9f+16)(f+1)^g - 3f^2 g^2 \nonumber \\
 &\quad&-\frac{1}{2}(3f^2+15f)-\frac{1}{2}(f^2+9f+16)\,,\nonumber \\
\end{eqnarray}
respectively. Inserting Eqs.~(\ref{2f8})-(\ref{2f10}) and $E_g=(f+1)^g-1$ into Eq.~(\ref{2f7}), we can arrive at the explicit expression for $r(f,g)$ as
\begin{widetext}
\begin{equation}\label{2f11}
r(f,g)= \frac{ 2f[(f + 1 ) ^g -1][(f + 1 )^g + 6g -1] - {f^2}{[(f + 1 )^g - 1]}^2  + 3{[( f + 1 )^g - 1]} ^2   + 8g[(f + 1 )^g( g - 1 ) + g + 1 ]}{ (f + 1 )^{ 2g + 1 }(f + 7)- 2f(f + 1 )^g[f + 8 - (f + 5)g + 6f{g^2}] - 14(f + 1 )^g + f[f + 8 - 2(f + 5)g - 4f{g^2}] +7    }\,.
\end{equation}
\end{widetext}
In Fig.~\ref{PeaCoe}, we report the exact result for $r(f,g)$ provided by Eq.~(\ref{2f11}). From Fig.~\ref{PeaCoe}, it is obvious that for $f=1,2$, $r(f,g)$ is positive;  for $f=3$, $r(f,g)$ equals zero; while for $f\geq 4$, $r(f,g)$ is negative.

Equation~(\ref{2f11}) shows that for very large $g$, we have
\begin{eqnarray}
r(f,g)&\simeq&\frac{{{{(f+1)}^{2g}}( - {f^2} + 2f + 3)}}{{{{(f+1)}^{2g}}({f^2} + 8f + 7)}}\nonumber \\
&=&-\frac{f-3}{f+7}\,,
\end{eqnarray}
which decreases with $f$. When $f=1$ and $f=2$, $r(f,g)$ is equal to $\frac{1}{4}$ and $\frac{1}{9}$, respectively. Thus, for $f=1$ and $f=2$, $U_g$ is assortative. When $f=3$, $r(f,g)$ is equal to 0, indicating that the network is uncorrelated. When $f \in [4,\infty)$, $r(f,g)$ is negative. Concretely, when $f$ increases from 4 to $\infty$, $r(f,g)$ decreases from $-\frac{1}{11}$ to $-1$, showing that $U_g$ is disassortative.

The phenomenon that the Pearson degree correlation coefficient $r(f,g)$ decreases with $f$ can be explained heuristically as follows. Note that there are $E_g=(f+1)^g-1$ edges in $U_g$, which means that for those $N_{g-1}=(f+1)^{g-1}$ old nodes having a degree higher than one, they have $2E_{g-1}+L(g)=(f+2)(f+1)^{g-1}-2$ neighboring nodes, among which $L(g)=f(f+1)^{g-1}$ neighbors are those newly generated nodes with a single degree. Thus, for large $g$, the fraction of neighbors with single degree is approximatively equal to $f/(f+2)$, which is an increasing function of $f$, meaning that in networks corresponding to larger $f$, the average degree of neighbors of old nodes is smaller.

\section{Laplacian eigenvalues and their corresponding eigenvectors}\label{Section3}

Although for general graphs, it is a challenge to determine their Laplacian eigenvalues and eigenvectors, as will be shown, for $U_g$ this problem can be settled.

\subsection{Eigenvalues}\label{Section31}

Let $\mathbf{A}_g=[A_{ij}]_{(f+1)^g \times (f+1)^g}$ denote the adjacency matrix of $U_g$, where $A_{ij} =A_{ji}=1$
if nodes $i$ and $j$ are adjacent, $A_{ij} =A_{ji}=0$ otherwise,
then the degree of node $i$ is $d_{i}= \sum_{j\in
U_g}A_{ij}$. Let $\mathbf{D}_g ={\rm diag} (d_{1},
d_{2},\ldots, d_{{{(f+1)^g}}})$ denote the diagonal degree
matrix of $U_g$, then the Laplacian matrix of $U_g$ is defined by
$\mathbf{L}_g=\mathbf{D}_g-\mathbf{A}_g$.

We first study the eigenvalues of $U_g$, leaving the
eigenvectors to Subsection~\ref{vectors}. By construction, it is easy to
see that $\textbf{A}_g$ and $\textbf{D}_g$ obey the following relations:
\begin{equation}\label{3f1}
\mathbf{A}_g = {\left( {\begin{array}{*{20}{c}}
\mathbf{A}_{g - 1}&\mathbf{I}_{g-1}&\mathbf{I}_{g-1}& \cdots &\mathbf{I}_{g-1}\\
\mathbf{I}_{g-1}&\textbf{0}&\textbf{0}& \cdots &\textbf{0}\\
\mathbf{I}_{g-1}&\textbf{0}&\textbf{0}& \cdots &\textbf{0}\\
 \vdots & \vdots & \vdots &\ddots& \vdots \\
\mathbf{I}_{g-1}&\textbf{0}&\textbf{0}& \cdots &\textbf{0}
\end{array}} \right)}
\end{equation}
and
\begin{equation}\label{3f2}
\mathbf{D}_g= {\left( {\begin{array}{*{20}{c}}
{{\mathbf{D}_{g - 1}} + f\mathbf{I}_{g-1}}&\textbf{0}&\textbf{0}& \cdots &\textbf{0}\\
\textbf{0}&\mathbf{I}_{g-1}&\textbf{0}& \cdots &\textbf{0}\\
\textbf{0}&\textbf{0}&\mathbf{I}_{g-1}& \cdots &\textbf{0}\\
 \vdots & \vdots & \vdots &\ddots& \vdots \\
\textbf{0}&\textbf{0}&\textbf{0}& \cdots &\mathbf{I}_{g-1}
\end{array}} \right)}\,
\end{equation}
in which each block is a ${(f + 1)^{g - 1}} \times {(f + 1)^{g - 1}}$ matrix and $\mathbf{I}_{g-1}$ is the ${(f + 1)^{g - 1}} \times {(f + 1)^{g - 1}}$ identity matrix. Thus, the Laplacian matrix of $U_g$ satisfies the following recursive relation:
\begin{small}
\begin{eqnarray}\label{3f3}
\mathbf{L}_g &= &\mathbf{D}_g - \mathbf{A}_g \nonumber \\
&= & {\left( {\begin{array}{*{20}{c}}
{{\mathbf{L}_{g - 1}} + f\mathbf{I}_{g-1}}&-\mathbf{I}_{g-1}&-\mathbf{I}_{g-1}& \cdots &-\mathbf{I}_{g-1}\\
-\mathbf{I}_{g-1}&\mathbf{I}_{g-1}&\textbf{0}& \cdots &\textbf{0}\\
-\mathbf{I}_{g-1}&\textbf{0}&\mathbf{I}_{g-1}& \cdots &\textbf{0}\\
 \vdots & \vdots & \vdots &\ddots& \vdots \\
-\mathbf{I}_{g-1}&\textbf{0}&\textbf{0}& \cdots &\mathbf{I}_{g-1}
\end{array}} \right)}
\end{eqnarray}
\end{small}

Obviously, the problem of determining Laplacian eigenvalues of $U_g$ is equivalent to finding the roots of characteristic polynomial $P_g(\lambda)$ of $\mathbf{L}_g$. To find the eigenvalues of $\mathbf{L}_g$, we just need to determine the roots of  $P_g(\lambda)$, which reads:
\begin{widetext}
\begin{eqnarray}\label{3f4}
P_t(\lambda)&=&{\rm det}(\lambda\textbf{I}_g-\textbf{L}_g)\nonumber\\
&=&{\rm
det}\left(\begin{array}{ccccc}(\lambda-f)\mathbf{I}_{g-1}-\textbf{L}_{g-1} &
\mathbf{I}_{g-1}& \mathbf{I}_{g-1} &\cdots& \mathbf{I}_{g-1}
\\\mathbf{I}_{g-1} & (\lambda-1)\mathbf{I}_{g-1} & {\textbf{0}}&\cdots& {\textbf{0}}
\\\mathbf{I}_{g-1} & {\textbf{0}}& (\lambda-1)\mathbf{I}_{g-1} &\cdots& {\textbf{0}}
\\\vdots & \vdots & \vdots & \ddots  &\vdots
\\\mathbf{I}_{g-1} & {\textbf{0}}& {\textbf{0}}&\cdots& (\lambda-1)\mathbf{I}_{g-1}
\end{array}\right)\nonumber
\\&=&\{{\rm det}[(\lambda-1)\mathbf{I}_{g-1}]\}^f\cdot{\rm det}\left(\begin{array}{ccccc}(\lambda-f)\mathbf{I}_{g-1}-\textbf{L}_{g-1}
& \mathbf{I}_{g-1}& \mathbf{I}_{g-1} &\cdots& \mathbf{I}_{g-1}
\\\frac{1}{\lambda-1}\mathbf{I}_{g-1} & \mathbf{I}_{g-1} & {\textbf{0}}&\cdots& {\textbf{0}}
\\\frac{1}{\lambda-1}\mathbf{I}_{g-1} & {\textbf{0}}& \mathbf{I}_{g-1} &\cdots& {\textbf{0}}
\\\vdots & \vdots & \vdots & \ddots  &\vdots
\\\frac{1}{\lambda-1}\mathbf{I}_{g-1} & {\textbf{0}}& {\textbf{0}}&\cdots& \mathbf{I}_{g-1}
\end{array}\right)\nonumber
\\&=&\{ {\rm det}[(\lambda-1)\mathbf{I}_{g-1}]\}^f\cdot{\rm det}\left(\begin{array}{ccccc}(\lambda-f-\frac{f}{\lambda-1})\mathbf{I}_{g-1}-\textbf{L}_{g-1}
& \textbf{0}& \textbf{0} &\cdots& \textbf{0}
\\\frac{1}{\lambda-1}\mathbf{I}_{g-1} & \mathbf{I}_{g-1} & {\textbf{0}}&\cdots& {\textbf{0}}
\\\frac{1}{\lambda-1}\mathbf{I}_{g-1} & {\textbf{0}}& \mathbf{I}_{g-1} &\cdots& {\textbf{0}}
\\\vdots & \vdots & \vdots & \ddots  &\vdots
\\\frac{1}{\lambda-1}\mathbf{I}_{g-1} & {\textbf{0}}& {\textbf{0}}&\cdots& \mathbf{I}_{g-1}
\end{array}\right)\,,
\end{eqnarray}
\end{widetext}
where we have used the elementary operations of matrix. Based on the results in~\cite{Si00}, $P_g(\lambda)$ can be expressed as
\begin{small}
\begin{equation}\label{3f5}
P_g(\lambda ) = {\{ \det[ (\lambda  - 1)\textbf{I}_{g-1}]\} ^f}\det \left [\left(\lambda  - f - \frac{f}{{\lambda  - 1}}\right)\textbf{I}_{g-1} - \textbf{L}_{g-1}\right]\,.
\end{equation}
\end{small}
Hence, $P_g(\lambda)$ can be further recast recursively as
\begin{equation}\label{3f6}
P_g(\lambda ) = {(\lambda  - 1)^{f{{(f + 1)}^{g - 1}}}}{P_{g - 1}}(\varphi (\lambda ))
\end{equation}
where $\varphi(\lambda)=\lambda-f-\frac{f}{\lambda-1}$. This recursion relation
provided in Eq.~(\ref{3f6}) is very useful for determining the eigenvalues and eigenvectors of the Laplacian matrix for $U_g$. Note that $P_{g - 1}(\lambda)$ is a monic polynomial of degree $(f+1)^{g-1}$, then the exponent of $\frac{f}{\lambda-1}$ in $P_{g-1}(\varphi(\lambda))$ is $(f+1)^{g-1}$, and the exponent of factor $(\lambda-1)$ in ${P_g}(\lambda )$ is
\begin{equation}\label{3f7}
f{(f + 1)^{g-1}} - {(f + 1)^{g - 1}} = (f - 1){(f + 1)^{g - 1}}\,.
\end{equation}
Therefore, $U_g$ has Laplacian eigenvalue 1 with  multiplicity $(f - 1){(f + 1)^{g - 1}}$.

It is evident that $U_g$ has $(f+1)^g$ Laplacian eigenvalues, denoted by $\lambda_1^g, \lambda_2^g, \ldots, \lambda_{(f+1)^g}^g$, the set of which is represented by $\Lambda_g$, i.e., $\Lambda_g = \{ \lambda_1^g, \lambda_2^g,\ldots, \lambda_{(f + 1)^g}^g \}$. In addition, without loss of generality, we assume that $\lambda_1 ^g \le \lambda_2^g \le \ldots \le \lambda_{(f + 1)^g}^g$. On the basis of above analysis, $\Lambda_g$ can be divided into two subsets $\Lambda_g^{(1)}$ and  $\Lambda_g^{(2)}$, such as $\Lambda_g=\Lambda_g^{(1)} \cup \Lambda_g^{(2)}$. $\Lambda_g^{(1)}$ contains all eigenvalues equal to 1, while $\Lambda_g^{(2)}$ includes the remain eigenvalues. Thus,
\begin{equation}\label{3f8}
\Lambda_g^{(1)}= \underbrace {\{ 1,1,1,\ldots,1,1\} }_{(f - 1){{(f + 1)}^{g - 1}}}\,,
\end{equation}
where the distinctness of elements is neglected.

The remaining $2(f+1)^{g-1}$ eigenvalues belonging to $\Lambda_g^{(2)}$ are determined by $P_{g-1}(\varphi(\lambda))=0$. Let the $2(f+1)^{g-1}$ eigenvalues be $\tilde{\lambda}^g_1,
\tilde{\lambda}^g_2,\ldots,\tilde{\lambda}^g_{2(f+1)^{f-1}}$, respectively. That is, $\Lambda_g^{(2)} = \{ \tilde \lambda _1^g,\tilde \lambda _2^g,\ldots,\tilde \lambda _{2{{(f + 1)}^{g - 1}}}^g\}$. For convenience, we assume that $\tilde \lambda _1^g \le \tilde \lambda _2^g \le \ldots \le \tilde \lambda _{2{{(f + 1)}^{g - 1}}}^g$. Equation~(\ref{3f6}) shows that for any element in $\Lambda_{g-1}$, say $\lambda_{i}^{g-1} \in \Lambda_{g-1}$, both solutions of
$\lambda-f-\frac{f}{\lambda-1}=\lambda_{i}^{g-1}$ are in $\Lambda_g^{(2)}$. It is clear that $\lambda-f-\frac{f}{\lambda-1}=\lambda_{i}^{g-1}$ is equivalent to
\begin{equation}\label{3f10}
{\lambda ^2} - (\lambda _i^{g - 1} + f + 1)\lambda  + \lambda _i^{g - 1} = 0\,,
\end{equation}
the two roots of which are denoted, respectively, by $\tilde{\lambda}_{i}^{g}$ and
$\tilde{\lambda}_{i+(f+1)^{g-1}}^{g}$, since these notations give a natural
increasing order of the eigenvalues of $U_g$, as will be shown below.

Solving the quadratic equation provided by Eq.~(\ref{3f10}), we obtain the two roots  to be $\tilde{\lambda}_{i}^{g}=r_1(\lambda_{i}^{g-1})$ and
$\tilde{\lambda}_{i+(f+1)^{g-1}}^{g}=r_2(\lambda_{i}^{g-1})$, where
$r_1(\lambda_{i}^{g-1})$ and $r_2(\lambda_{i}^{g-1})$ are
\begin{small}
\begin{equation}\label{3f11}
r_1(\lambda _i^{g - 1}) = \frac{1}{2}\left(\lambda _i^{g - 1} + f + 1 - \sqrt {{{(\lambda _i^{g - 1} + f + 1)}^2} - 4\lambda _i^{g - 1}}\right)
\end{equation}
\end{small}
and
\begin{small}
\begin{equation}\label{3f12}
r_2(\lambda _i^{g - 1}) = \frac{1}{2}\left(\lambda _i^{g - 1} + f + 1 + \sqrt {{{(\lambda _i^{g - 1} + f + 1)}^2} - 4\lambda _i^{g - 1}}\right)\,,
\end{equation}
\end{small}
respectively. Thus, in this way each eigenvalue $\lambda_i^{g - 1}$ in $\Lambda_{g-1}$ gives rise to two new eigenvalues in $\Lambda_g^{(2)}$.
Inserting each Laplacian eigenvalue of $U_{g - 1}$ into Eqs.~(\ref{3f11}) and~(\ref{3f12}) generates all the elements of $\Lambda_g^{(2)}$. Considering the initial value ${\Lambda_0} = \{ 0\} $, by recursively applying Eqs.~(\ref{3f11}) and~(\ref{3f12}), the Laplacian eigenvalues of $U_g$ can be fully determined.

It is easy to prove that the two roots, ${r_1}(\lambda _i^{g - 1})$ and ${r_2}(\lambda _i^{g - 1})$, of Eq.~(\ref{3f10}) monotonously increase with $\lambda _i^{g - 1}$ and both lie in intervals $[0,1)$ and $(1, +\infty)$, respectively. Thus, for any eigenvalue in $\lambda _i^{g - 1} \in \Lambda_{g - 1}$, ${r_1}(\lambda _i^{g - 1}) < 1 < {r_2}(\lambda _i^{g - 1})$ always holds. In addition,  the following conclusion can be reached based on simple argument. Assuming that ${E_{g - 1}} = \{ \lambda _1^{g - 1},\lambda _2^{g - 1},...,\lambda _{{{(f + 1)}^{g - 1}}}^{g - 1}\}$, then $\Lambda_g^{(2)}$ can be generated via Eqs.~(\ref{3f11}) and ~(\ref{3f12}), that is,  $\Lambda_g^{(2)} = \{ \tilde \lambda _1^g,\tilde \lambda _2^g,...,\tilde \lambda _{2{{(f + 1)}^g}}^g\} $ satisfying
$\tilde \lambda _1^g \le  \lambda _2^g \cdots \le \tilde \lambda _{{{(f + 1)}^{g - 1}}}^g < 1 < \tilde \lambda _{{{(f + 1)}^{g - 1}} + 1}^g \le \tilde \lambda _{{{(f + 1)}^{g - 1}} + 2}^g \cdots \le \tilde \lambda _{2{{(f + 1)}^{g - 1}}}^g$. Recall that $\Lambda_g^{(1)}$ contains
$(f - 1){(f + 1)^{g - 1}}$ elements 1, we now have gotten the whole set of Laplacian eigenvalues for $U_g$ to be $\Lambda_g = \Lambda_g^{(1)} \cup \Lambda_g^{(2)}$.

\begin{figure}[h]
\centering
\includegraphics[width=1\linewidth,trim=0 0 0 0]{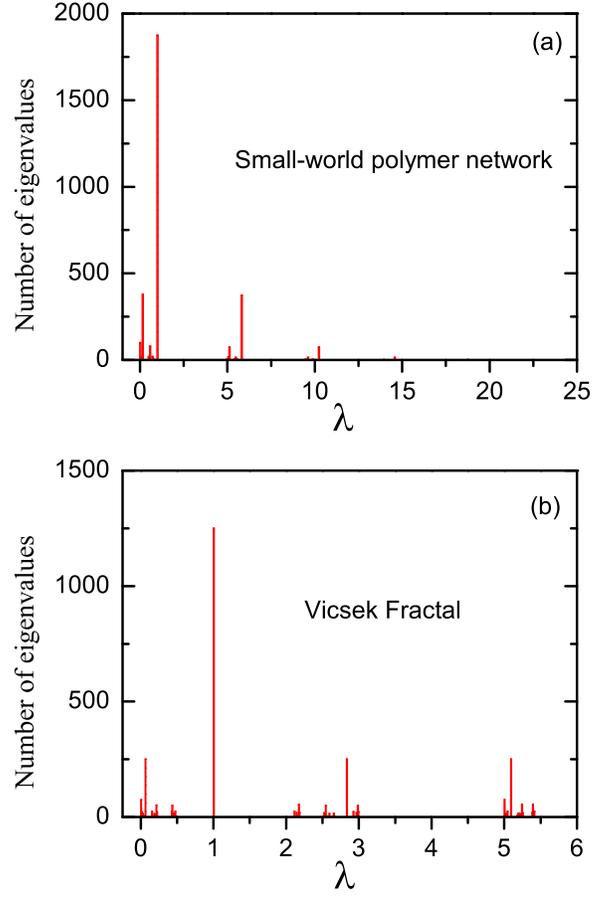}
\caption{Number of distinct eigenvalues for a small-world polymer network (a) and its corresponding Vicsek fracal, with $f=4$ and $g=5$ for both networks.}\label{EigenV}
\end{figure}

In order to see the distribution of the Laplacian eigenvalues for $U_g$. We use Eqs.~(\ref{3f11}) and~(\ref{3f12}) to determine the eigenvalues of a specifical network corresponding to $f=4$ and $g=5$. In addition, by diagonalizing the associated Laplacian matrix, we also compute numerically the eigenvalues and their multiplicities, which are in complete agreement with those analytical results, confirming that the theoretic approach is valid. In Fig.~\ref{EigenV}(a), we display as a histogram, for the result of the network corresponding to $f=4$ and $g=5$, thus having a size $N_5=3125$. Furthermore, we also present in Fig.~\ref{EigenV}(b) the histogram for the corresponding Vicsek fractals with $f=4$ and $g=5$.

By comparing Figs.~\ref{EigenV}(a) and (b), we can see that number of distinct eigenvalues in the small-world network is much less than its corresponding Vicsek fractal. Note that in $U_g$, the distinct degree values for nodes are $g+1$, while for corresponding Vicsek fractals, the degree values are 3 (all node have degree 1, 2, or $f$). The reasons for the interesting phenomenon that Vicsek fractals display a larger heterogeneity in the Laplacian spectrum but a far smaller heterogeneity in the degree values deserves further study in the future. In addition to the number of dissimilar eigenvalues, the difference of eigenvalues are also obvious for these two networks. For instance, the maximum eigenvalue, $\lambda_{\max}^g$, of the small-world polymer network is substantially higher than that of the Viscek fractal. As we will show, these differences of Laplacian spectra between the two networks will lead to different behaviors for various dynamics taking place on them.

\subsection{Eigenvectors}\label{vectors}

Analogous to the eigenvalues, the eigenvectors of $\textbf{L}_g$
can also be derived directly from those of $\textbf{L}_{g-1}$. Assume that
$\lambda$ is an eigenvalue of Laplacian matrix for $U_g$, the
corresponding eigenvector of which is $\textbf{\emph{v}} \in
\textbf{R}^{(f+1)^g}$, where $\textbf{R}^{(f+1)^g}$ is the
$(f+1)^g$-dimensional vector space. Then the eigenvector $\textbf{\emph{v}}$ can be determined by solving equation
($\lambda\, \textbf{I}_g -\textbf{L}_g)\textbf{\emph{v}}=0$. We distinguish two cases:
$\lambda \in \Lambda_g^{(1)}$ and $\lambda \in \Lambda_g^{(2)}$, which will be
separately treated as follows.

For the case of $\lambda \in \Lambda_g^{(1)}$, in which all $\lambda=1$,
equation ($\lambda\, \textbf{I}_g
-\textbf{L}_g)\textbf{\emph{v}}=0$ becomes
\begin{small}
\begin{equation}\label{vector11}
\left(\begin{array}{ccccc}(1-f)\textbf{I}_{g-1}-\textbf{L}_{g-1} &
\textbf{I}{g-1}& \textbf{I}{g-1} &\cdots& \textbf{I}{g-1}
\\\textbf{I}_{g-1} &\textbf{0}& {\textbf{0}}&\cdots& {\textbf{0}}
\\\textbf{I}_{g-1} & {\textbf{0}}& \textbf{0}&\cdots& {\textbf{0}}
\\\vdots & \vdots & \vdots & \ddots  &\vdots
\\\textbf{I}_{g-1} & {\textbf{0}}& {\textbf{0}}&\cdots& \textbf{0}
\end{array}\right)\left(\begin{array}{c}\textbf{\emph{v}}_1\\\textbf{\emph{v}}_2\\\textbf{\emph{v}}_3 \\\vdots\\ \textbf{\emph{v}}_{f+1}
\end{array}\right)=0,
\end{equation}
\end{small}
where vector $\textbf{\emph{v}}_i$ ($1\le i \le f+1$) are components
of $\textbf{\emph{v}}$. Equation~(\ref{vector11}) leads to the
following equations:
\begin{eqnarray}
\textbf{\emph{v}}_1=\textbf{0},\label{vector12}
\\\textbf{\emph{v}}_2+\textbf{\emph{v}}_3+\dots+\textbf{\emph{v}}_{f+1}=\textbf{0}.\label{vector13}
\end{eqnarray}
In Eq.~(\ref{vector12}), $\textbf{\emph{v}}_1$ is a zero vector. Let $\textbf{\emph{v}}_i=(\textbf{\emph{v}}_{i,1},\textbf{\emph{v}}_{i,2},\ldots,\textbf{\emph{v}}_{i,(f+1)^g})^\top$, then, Eq.~(\ref{vector13}) is equivalent to the following equations:

\begin{small}
\[ \left \{
\begin{array}{ccccccccc}
\textbf{\emph{v}}_{2,1}&+&\textbf{\emph{v}}_{3,1}&+&\dots
&+&\textbf{\emph{v}}_{f+1,1}=\textbf{0}
\\\textbf{\emph{v}}_{2,2}&+&\textbf{\emph{v}}_{3,2}&+&\dots
&+&\textbf{\emph{v}}_{f+1,2}=\textbf{0}
\\ \vdots & \vdots  &\vdots  & \vdots  &\  &\vdots &\vdots
\\\textbf{\emph{v}}_{2,(f+1)^{g-1}}&+&\textbf{\emph{v}}_{3,(f+1)^{g-1}}&+&\dots
&+&\textbf{\emph{v}}_{f+1,(f+1)^{g-1}}=\textbf{0}
\end{array} \right.
\]
\end{small}

The set of all solutions to any of the above equations consists of
vectors of the following form
\begin{widetext}
\begin{equation}\label{vector15}
\left(\begin{array}{ccc}\textbf{\emph{v}}_{2,j}
\\\textbf{\emph{v}}_{3,j}\\\textbf{\emph{v}}_{4,j}\\\vdots\\\textbf{\emph{v}}_{f+1,j} \end{array}\right)=
k_{1,j}
\left(\begin{array}{ccc}\textbf{-1}\\
\textbf{1}\\
\textbf{0} \\ \vdots \\
\textbf{0} \end{array}\right)+ k_{2,j}
\left(\begin{array}{ccc}\textbf{-1}\\
\textbf{0}\\
\textbf{1} \\ \vdots \\
\textbf{0} \end{array}\right)+\dots+k_{m-1,j}
\left(\begin{array}{ccc}\textbf{-1}\\
\textbf{0}\\
\textbf{0} \\ \vdots \\
\textbf{1} \end{array}\right),
\end{equation}
where $k_{1,j}$ , $k_{2,j}$ , $\dots$ , $k_{f-1,j}$ are arbitrary
real numbers. In Eq.~(\ref{vector15}), the solutions for all the
vectors $\textbf{\emph{v}}_i$ ($2\le i \le f+1$) can be rewritten as
\begin{equation}\label{vector16}
\left(\begin{array}{c}\textbf{\emph{v}}_2^\top\\\textbf{\emph{v}}_3^\top\\\textbf{\emph{v}}_4^\top
\\\vdots\\ \textbf{\emph{v}}_{f+1}^\top
\end{array}\right)=
\left(\begin{array}{cccc}\textbf{-1} & \textbf{-1}& \cdots&
\textbf{-1}
\\\textbf{1} & {\textbf{0}}&\cdots& {\textbf{0}}
\\\textbf{0} & {\textbf{1}}& \cdots& {\textbf{0}}
\\\vdots & \vdots &  \  &\vdots
\\\textbf{0} & {\textbf{0}}& \cdots& \textbf{1}
\end{array}\right)\\
\\
\left(\begin{array}{cccc}k_{1,1} & k_{1,2}& \cdots&
k_{1,(f+1)^{g-1}}
\\k_{2,1} &k_{2,2} & \cdots & k_{2,(f+1)^{g-1}}
\\k_{3,1} & k_{3,2}& \cdots& k_{3,(f+1)^{g-1}}
\\\vdots  & \vdots & \  &\vdots
\\k_{f-1,1}  & k_{f-1,2} &\cdots& k_{f-1,(f+1)^{g-1}}
\end{array}\right),
\end{equation}
\end{widetext}
where $k_{i,j}$ \big($1\le i\le f-1$; $1\le j\le (f+1)^{g-1}$\big)
are arbitrary real numbers. Using Eq.~(\ref{vector16}), we
can obtain the eigenvector $\textbf{\emph{v}}$ associated with the
eigenvalue 1. Furthermore, we can easily check that the
dimension of the eigenspace of matrix $\textbf{L}_g$ corresponding to
eigenvalue 1 is $(f-1)(f+1)^{g-1}$.

We proceed to address the case of $\lambda \in \Lambda_g^{(2)}$. For this case, equation ($\lambda\, \textbf{I}_g -\textbf{L}_g)\textbf{\emph{v}}=0$
can be rewritten as
\begin{widetext}
\begin{equation}\label{vector01}
\left(\begin{array}{ccccc}(\lambda-f)\textbf{I}_{g-1}-\textbf{L}_{g-1} &
\textbf{I}_{g-1}& \textbf{I}_{g-1} &\cdots& \textbf{I}_{g-1}
\\\textbf{I}_{g-1} &(\lambda-1)\textbf{I}_{g-1}& {\textbf{0}}&\cdots& {\textbf{0}}
\\\textbf{I}_{g-1} & {\textbf{0}}& (\lambda-1)\textbf{I}_{g-1}&\cdots& {\textbf{0}}
\\\vdots & \vdots & \vdots & \ddots  &\vdots
\\\textbf{I}_{g-1} & {\textbf{0}}& {\textbf{0}}&\cdots& (\lambda-1)\textbf{I}_{g-1}
\end{array}\right)\left(\begin{array}{c}\textbf{\emph{v}}_1\\\textbf{\emph{v}}_2\\\textbf{\emph{v}}_3 \\\vdots\\ \textbf{\emph{v}}_{f+1}
\end{array}\right)=0,
\end{equation}
\end{widetext}
where vector $\textbf{\emph{v}}_i$ ($1\le i \le f+1$) are components
of $\textbf{\emph{v}}$. Equation~(\ref{vector01}) leads to the
following equations:
\begin{eqnarray}
\big[(\lambda-f)\textbf{I}_{g-1}-\textbf{L}_{g-1}\big]\textbf{\emph{v}}_1+\textbf{\emph{v}}_2+\dots+\textbf{\emph{v}}_{f+1}=\textbf{0},\label{vector02}
\\ \textbf{\emph{v}}_1+(\lambda-1)\textbf{\emph{v}}_i=\textbf{0}\ \ \  (2\le i \le f+1).\label{vector03}
\end{eqnarray}
Resolving Eq.~(\ref{vector03}) yields
\begin{eqnarray}\label{vector04}
\textbf{\emph{v}}_i=-\frac{1}{\lambda-1}\textbf{\emph{v}}_1 \ \
(2\le i \le f+1).
\end{eqnarray}
Inserting Eq.~(\ref{vector04}) into Eq.~(\ref{vector02}) results in
\begin{eqnarray}\label{vector05}
\left[\left(\lambda-f-\frac{f}{\lambda-1}\right)\textbf{I}_{g-1}-\textbf{L}_{g-1}\right]\textbf{\emph{v}}_1=0,
\end{eqnarray}
which indicates that $\textbf{\emph{v}}_1$ is the solution of
Eq.~(\ref{vector02}) while $\textbf{\emph{v}}_i$ ($2\le i\le f+1$)
are completely determined by $\textbf{\emph{v}}_1$ via
Eq.~(\ref{vector04}). As demonstrated in Eq.~(\ref{3f6}), if $\lambda$ is an
eigenvalue of $\textbf{L}_g$, then
$\varphi(\lambda)=\lambda-f-\frac{f}{\lambda-1}$ is an eigenvalue of
$\textbf{L}_{g-1}$. Thus, Eqs.~(\ref{vector05}) and~(\ref{3f6}) implies that $\textbf{\emph{v}}_1$ is an eigenvector of $\textbf{L}_{g-1}$ corresponding to eigenvalue $\lambda-f-\frac{f}{\lambda-1}$, while
\begin{equation}
\textbf{\emph{v}}=\left(\begin{array}{ccc}\textbf{\emph{v}}_1
\\\textbf{\emph{v}}_2\\\textbf{\emph{v}}_3\\\vdots\\\textbf{\emph{v}}_{f+1} \end{array}\right)=
\left(\begin{array}{ccc}\textbf{\emph{v}}_1\\
-\frac{1}{\lambda-1}\textbf{\emph{v}}_1\\
-\frac{1}{\lambda-1}\textbf{\emph{v}}_1 \\ \vdots \\
-\frac{1}{\lambda-1}\textbf{\emph{v}}_1\end{array}\right)
\end{equation}
is an eigenvector of $\textbf{L}_g$ associated with eigenvalue
$\lambda$.

Since for the initial graph $U_0$, its Laplacian matrix
$\textbf{L}_0$ has only one eigenvalue 0 with corresponding
eigenvector $(1)^\top$; by recursively applying
the above process, we can obtain all the eigenvectors corresponding
to $\lambda \in \Lambda_g^{(2)}$.

In this way, we have completely determined all eigenvalues and their corresponding eigenvectors of $U_g$. In the following text, we will use these obtained results, especially those for eigenvalues, to study some dynamical processes taking places in $U_g$, including random walks with a trap, relaxation dynamics in the GGS framework, and depolarization of fluorescence by F\"{o}ster quasiresonant energy transfer.

\section{Trapping process} \label{Section7}

In this section, we study trapping problem in the small-world polymer networks. The trapping problem is a particular kind of random walks with a trap fixed at a position, absorbing all particles visiting it. In the process of random walks, at each time step, the
particle (walker), starting from its current location, moves to any
of its nearest neighbors with equal probability. One of the primary
quantities related to trapping problem is trapping time (TT)~\cite{Re01}. The TT for a node is defined as the mean first-passage time (MFPT) for a particle starting from the node to the trap. Let $F_{i,j}(g)$ denote the MFPT from node $i$ to node $j$. Below we will focus on two cases of trapping problem. In the first case, the trap is fixed on the central node, while in the other case, the trap is uniformly distributed over the whole networks.

\subsection{Trapping with a trap fixed on the central node}\label{Section71}

We first consider the case of trapping in $U_{g}$ with the perfect trap being located at the central hub node $h_{g}$. In this case, the quantity we are concerned with is the average trapping time (ATT), $F_{h}(g)$, which is the average of $F_{i,h_g}(g)$ over all possible starting points in $U_{g}$. That is,
\begin{equation}\label{8f1}
F_{h}(g)=\frac{1}{N_g}\sum_{i=1}^{N_g} F_{i,h_g}(g)\,.
\end{equation}
We next study analytically $F_{h}(g)$ by using the second construction method of the networks, showing how $F_{h}(g)$ changes with the network size $N_g$.

Let $F_{\rm sum}(g)$ denote the sum term on the rhs of Eq.~(\ref{8f1}), i.e.,
\begin{eqnarray}\label{8f2}
F_{\rm sum}(g) = \sum_{i \in U_g}F_{i,h_g}(g)\,.
\end{eqnarray}
Then,
\begin{equation}\label{8f3}
F_{h}(g)=\frac{F_{\rm sum}(g)}{N_g}\,.
\end{equation}
Thus, we reduce the problem of determining $F_{h}(g)$ to evaluating $F_{\rm sum}(g)$.
To find $F_{\rm sum}(g)$, we should determine some intermediary quantities.
First, for all $g \geq 0$, $F_{h_g,h_g}(g)=0$. On the other hand, according to the previous results obtained by various techniques~\cite{NoRi04,LiWuZh10}, we have
\begin{equation}\label{8f4}
F_{h_g^{(i)},h_g}(g) =F_{h_g^{(i)},h_g^{(0)}}(g)= 2N_{g-1} - 1 = 2(f + 1)^{g-1} - 1
\end{equation}
for all $1\leq i \leq f$. Then, from the second construction of the networks, we obtain
\begin{eqnarray}\label{8f5}
&\quad& F_{\rm sum}(g) \nonumber\\
&=& \sum\limits_{i \in U_{g - 1}^{(0)}} {{F_{i,h_g}}(g)}  + \sum\limits_{j = 1}^f {\sum\limits_{i \in U_{g - 1}^{(j)}} {\left[F_{i,h_g^{(j)}}(g) + F_{h_g^{(i)},h_g^{(0)}}(g)\right]}} \nonumber\\
 &=&  F_{\rm sum}(g-1) + f[ F_{\rm sum}(g-1) + N_{g-1}(2N_{g-1}- 1)]\nonumber\\
 &=& (f + 1)F_{\rm sum}(g-1)+ f(f + 1)^{g-1}[ 2(f + 1)^{g-1} - 1]\,.\nonumber\\
\end{eqnarray}
Considering $F_{\rm sum}(0)=0$, Eq.~(\ref{8f5}) is solved to yield
\begin{equation}\label{8f6}
F_{\rm sum}(g) =2(f+1)^{2g-1}-(f+1)^{g-1}(fg + 2 )\,.
\end{equation}

Substituting Eq.~(\ref{8f6}) into Eq.~(\ref{8f3}), we arrive at the closed-form expression of $F_{h}(g)$ as
\begin{equation}\label{8f7}
F_{h}(g) = 2(f+1)^{g-1} - \frac{fg + 2}{f+1}\,.
\end{equation}
We next show how to represent $F_{h}(g)$ in terms  of the network size $N_g$, with a goal to obtain the relation between these two quantities. Recalling Eq.~(\ref{1f1}), we
have $g=\ln N_g / \ln (f+1)$, which enables to write $F_{h}(g)$ in the following
form:
\begin{equation}\label{8f8}
F_{h}(g) = \frac{2{N_g}}{f+1} - \frac{f\ln N_g}{(f+1)\ln (f+1)} - \frac{2}{f+1}\,.
\end{equation}
Equation~(\ref{8f8}) provides an explicit dependence relation
of $F_{h}(g)$ on $N_g$ and parameter $f$. For a sufficiently large
system, i.e., $N_g\rightarrow \infty$, the dominating term of $F_{h}(g)$ is
\begin{equation}\label{8f9}
F_{h}(g) \simeq \frac{2{N_g}}{f+1}\,,
\end{equation}
which increases linearly with the system size. This linear scaling of ATT on the network size is in sharp contrast to the superlinear scaling of ATT in Vicsek fractals with the central node as the trap~\cite{WuLiZhCh12,LiZh13}.

\subsection{Trapping with the trap uniformly distributed} \label{Section72}

In Subsection~\ref{Section71}, we have discussed the trapping problem in $U_{g}$ with an immobile trap positioned at the central node. Here we study another case of trapping problem in $U_{g}$ with the trap uniformly distributed over the whole networks. In this case, we are concerned with the quantity $F_g$ defined as the average of MFPT $F_{ij}(g)$ over all pairs of source point $i$ and target point $j$ in the networks:
\begin{eqnarray}\label{E00}
F_g=\frac{1}{(N_g)^2}\sum_{i=1}^{N_g}\sum_{j=1}^{N_g} F_{ij}(g)\,.
\end{eqnarray}
Let $F_{\rm tot}(g)$ denote the summation term on the rhs of  Eq.~(\ref{E00}):
\begin{eqnarray}\label{E01}
F_{\rm tot}(g)=\sum_{i=1}^{N_g}\sum_{j=1}^{N_g} F_{ij}(g)\,.
\end{eqnarray}
Then,
\begin{eqnarray}\label{E000}
F_g=\frac{F_{\rm tot}(g)}{(N_g)^2}\,,
\end{eqnarray}
which is actually the ATT when the trap is uniformly distributed. Notice that the quantity $F_g$ involves a double average: the first one is over all the source points to a given trap, the second one is the average of the first one.

In order to compute $F_g$, we use the relation governing resistance distance and MFPTs between two nodes in a connected graph~\cite{ChRaRuSm89,Te91}. For this purpose, we look on $U_{g}$ as an electrical network~\cite{DoSn84} by considering each edge in $U_{g}$ to be a unit resistor~\cite{KlRa93}. Let $R_{ij}(g)$ be the effective resistance between two nodes $i$ and $j$ in the electrical network corresponding to $U_{g}$. Then, the following exact relation
\begin{eqnarray}\label{H00}
F_{ij}(g)+F_{ji}(g)=2E_g\, R_{ij}(g)\,
\end{eqnarray}
holds~\cite{ChRaRuSm89,Te91}, and Eq.~(\ref{E01}) can be recast as
\begin{eqnarray}\label{H01}
F_{\rm tot}(g)=E_g\sum_{i=1}^{N_g}\sum_{j=1}^{N_g}R_{ij}(g)\,.
\end{eqnarray}
Applying the previous
results~\cite{GuMo96,ZhKlLu96}, the sum term of effective resistance between all pairs of nodes in $U_{g}$ can be evaluated as
\begin{equation}\label{Hitting08}
\sum_{i=1}^{N_g}\sum_{j=1}^{N_g}R_{ij}(g)=2N_g\,\sum_{i=2}^{N_g}\frac{1}{\lambda_i^g}\,.
\end{equation}
Then, Eq.~(\ref{E00}) becomes
\begin{equation}\label{7f5}
F _g=2\,\sum_{i=2}^{N_g}\frac{1}{\lambda_i^g}\,.
\end{equation}

Having expressing $F_g$ in terms of the sum of the reciprocal
of all nonzero Laplacian eigenvalues for $U_g$, the next step is to find
this sum, denoted by $\Gamma_g$. By definition,
\begin{equation}\label{7f6}
\Gamma_g= \sum_{i = 2}^{N_g} \frac{1}{\lambda_i^g} = \sum_{\lambda _i^g \in \Lambda_g^{(1)}} {\frac{1}{\lambda_i^g}  + \sum_{{\tilde \lambda}_i^g \in \Lambda_g^{(2)}} \frac{1}{{{\tilde \lambda}_i^g}}}\,.
\end{equation}
Let $\Gamma_g^{(1)}$ and $\Gamma_g^{(2)}$ denote separately the two sums on the rhs of Eq.~(\ref{7f6}). Obviously,
\begin{eqnarray}\label{7f7}
\Gamma_g^{(1)} =(f-1)(f+1)^{g-1}\,.
\end{eqnarray}
And $\Gamma_g^{(2)}$ can also be calculated as
\begin{eqnarray}\label{7f8}
\Gamma_g^{(2)} &=& \sum\limits_{i = 2}^{2{{(f + 1)}^{g - 1}}} {\frac{1}{{\tilde \lambda  _i^g}}} \nonumber\\
 &=& \sum\limits_{i = 2}^{{{(f + 1)}^{g - 1}}} {\left( {\frac{1}{{\tilde \lambda _i^g}} + \frac{1}{{\tilde \lambda _{i + {{(f + 1)}^{g - 1}}}^g}}} \right)}  + \frac{1}{{\tilde \lambda _{1 + {{(f + 1)}^{g - 1}}}^g}}\nonumber\\
 &=& \sum\limits_{i = 2}^{{{(f + 1)}^{g - 1}}} {\frac{{\tilde \lambda _i^g + \tilde \lambda _{i + {{(f + 1)}^{g - 1}}}^g}}{{\tilde \lambda _i^g\tilde \lambda _{i + {{(f + 1)}^{g - 1}}}^g}}}  + \frac{1}{{\tilde \lambda _{1 + {{(f + 1)}^{g - 1}}}^g}}\,.
\end{eqnarray}
Because ${\tilde \lambda} _i^g$ and ${\tilde \lambda}^g_{i+(f + 1)^{g - 1}}$ are two roots of the quadratic equation given by Eq.~(\ref{3f10}), using Vieta's formulas, we have $\tilde \lambda _i^g + \tilde \lambda _{i + {{(f + 1)}^{g - 1}}}^g = \lambda _i^{g - 1} + f + 1$ and $\tilde \lambda _i^g \times  \tilde \lambda _{i + {{(f + 1)}^{g - 1}}}^g = \lambda _i^{g - 1}$. Furthermore, considering $\tilde \lambda _1^g=0$, so $\tilde \lambda _{1 + {{(f + 1)}^{g - 1}}}^g = m + 1$. Then Eq.~(\ref{7f8}) is reduced to
\begin{eqnarray}\label{7f9}
\Gamma_g^{(2)}&=& \sum\limits_{i = 2}^{{{(f + 1)}^{g - 1}}} {\frac{{\lambda _i^{g - 1} + f + 1}}{{\lambda _i^{g - 1}}}}  + \frac{1}{{f + 1}}\nonumber\\
&=& {(f + 1)^{g - 1}} - 1 + (f + 1)\sum\limits_{i = 2}^{{{(f + 1)}^{g - 1}}} {\frac{1}{{\lambda _i^{g - 1}}}}  + \frac{1}{{f + 1}}\nonumber\\
&=& {(f + 1)^{g - 1}} - 1 + (f + 1)T_{g - 1} + \frac{1}{{f + 1}}\,.
\end{eqnarray}

Note that $\Gamma_g^{(2)} = \Gamma_g-\Gamma_g^{(1)} =  \Gamma_g- (f - 1){(f + 1)^{g - 1}}$, applying this result into Eq.~(\ref{7f6}), one can reach the following recursive relation for $\Gamma_g$:
\begin{eqnarray}\label{7f10}
\Gamma_g=(f + 1)\Gamma_{g-1}+f{(f+1)^{g-1}}-\frac{f}{{f + 1}}\,.
\end{eqnarray}
With the initial situation $\Gamma_0=0$, Eq.~(\ref{7f10}) can be resolved to yield an explicit formula for $\Gamma_g$ as
\begin{eqnarray}\label{7f11}
\Gamma_g=(f+1)^{g-1}(fg - 1)+\frac{1}{f+1}\,.
\end{eqnarray}
Thus, the exact expression for $F_g$ is
\begin{eqnarray}\label{7f12}
F_g = 2(f+1)^{g-1}(fg - 1)+\frac{2}{f+1}\,,
\end{eqnarray}
which can be further represented as a function of network size $N_g$ as
\begin{eqnarray}\label{7f13}
F_g  =\frac{2f}{(f+1)\ln (f+1)} N_g \ln N_g -\frac{2}{f + 1}N_g+ \frac{2}{f + 1}\,.\nonumber \\
\end{eqnarray}
When the network size tends to infinity, i.e., $g \rightarrow \infty $, $F_g$ has the following dominant form
\begin{eqnarray}\label{7f14}
F_g \sim \frac{2f}{(f+1)\ln (f+1)} N_g \ln N_g\,,
\end{eqnarray}
a scaling also different from that previously found for Vicsek fractals~\cite{ZhWjZhZhGuWa10}, in which $F_g$ increases as a superlinear function of $N_g$.

\subsection{Result comparison and analysis} \label{Section73}

From above-obtained results given by Eqs.~(\ref{8f9}) and~(\ref{7f14}), it is easy to see that the dominating terms for $F_{h_g}(g)$ and $F _g$ behave differently. The former obeys $F_{h_g}(g) \sim N_g$, while the latter follows $ F_g \sim N_g \,\ln N_g $, greater than that of the former. This disparity indicates that in the family of treelike small-world polymer networks, the location of the trap has a strong influence on the trapping efficiency measured by ATT, which is in comparison with that for Vicsek fractals, where the effect of trap's location is negligible~\cite{WuLiZhCh12,LiZh13,ZhWjZhZhGuWa10}. In addition, the distinction between $F_{h_g}(g)$ and $F_g$ also shows that the leading scaling of ATT to a given node in $U_g$, e.g., the central node, might be not representative of the networks.

The dissimilar dominating scalings for $F_{h_g}(g)$ and $F_g$ in $U_g$ lie in the network structure and can be heuristically accounted for as follows. As shown in Fig.~\ref{self-simi}, $U_g$ consists of $f+1$ copies of $U_{g-1}$: one central replica, and $f$ peripheral duplicates. When the trap is positioned at the central hub node, the particle will visit at most one copy of $U_{g-1}$, i.e., a faction of $1/(f+1)$ among all nodes in $U_g$. Thus, the ATT $F_{h_g}(g)$ is small and grows linearly with network size, revealing a high trapping efficiency. In contrast, when the trap is located at another node, the particle should first visit the hub node, from which it continues to jump until being absorbed by the trap. So, the percentage of visited nodes is larger than that of the case when the trap is fixed at the hub. In particular, for the case that the trap is placed at a node farthest from the hub, the particle must visit all nodes of the networks before reaching the target. That is why the trapping process is less efficient when the trap is uniformly distributed.

The differences of behaviors of random walks in the small-world treelike polymer networks and Vicsek fractals are rooted in their underlying structures. For example, for trapping with a trap at the central node, the fact that the trapping efficiency of the former is higher than the latter can be understood as follows. for a walker in the small-world trees, as shown above, it will visit at most a faction of $\frac{1}{f+1}$ nodes before being trapped; while for trapping in Vicsek fractals, the walker may visit a larger fraction (greater than $\frac{1}{f+1}$) of nodes prior to being absorbed by the central trap node.

\section{Generalized Gaussian structures and relaxation patterns} \label{Section4}

In this section, we consider the relaxation dynamics of the treelike polymer networks in the framework of GGS~\cite{SoBl95,Sc98,BiKaBl00,KaBiBl00}, which is an extension of the classic Rouse model~\cite{Ro53}, developed for linear polymer chains and extended to more complex geometries.

\subsection{Brief introduction to GGS}

The theory of GGS was accounted for in detail in previous works~\cite{SoBl95,Sc98,BiKaBl00}, thus we give here only a brief introduction of the basic equation and main results related to the relaxation dynamics patterns.

A GGS consists of $N$ beads subject to the friction with friction constant $\zeta$, which are connected to each other by elastic springs with elasticity constant $K$. In the Langevin formalism, the dynamics of bead $m$ obey the following equation
\begin{equation}\label{4f1}
\zeta \frac{d \mathbf{R}_m(t)}{dt} + K\, \sum_{i=1}^N L_{mi}\mathbf{R}_m(t) = \mathbf{f}_m(t) + \mathbf{F}_m(t)\,.
\end{equation}
In Eq.~(\ref{4f1}), $\mathbf{R}_m(t)=(X_m(t),Y_{m}(t),Z_{m}(t))$ is the position vector of the $m$th bead at time $t$; $L_{mi}$ is the $mi$th entry of the Laplacian matrix $\mathbf{L}$ describing the topology of the GGS; $\mathbf{f}_m(t)$ is the thermal noise that is assumed to be Gaussian with zero mean value $\langle \mathbf{f}_m(t)\rangle$ and $\langle f_{m \alpha}(t)f_{m \beta}(t')\rangle=2k_B T\delta_{\alpha \beta}\delta(t-t')$, where $k_B$ is the Boltzmann constant, $T$ is the temperature, $\alpha$ and $\beta$ represent the $x$, $y$, and $z$ directions; $\mathbf{F}_m(t)$ is the external force acting on bead $m$.

We focus on the motion (drift and stretching) of the GGS under a constant external force $\mathbf{F} = F\Theta(t)\mathbf{e}_y$ (here $\Theta(t)$ is the Heaviside step function), switched on at $t=0$ and
acting on a single bead in the y direction. The displacement along the $y$ direction, $Y(t)$, after averaging both over the fluctuating forces $\mathbf{f}_m(t)$ and over all the beads in the GGS, reads~\cite{Sc98,BiKaBl00,KaBiBl00}
\begin{equation}\label{4f2}
\langle Y(t) \rangle = \frac{F t}{N \zeta} + \frac{F}{\sigma N \zeta} \sum_{i=2}^N \frac{1 -\exp(-\sigma \lambda_i t)}{\lambda_i}\,,
\end{equation}
where $\sigma=K/\zeta$ is the bond rate constant, and $\lambda_i$ is the eigenvalues of matrix $\mathbf{L}$ with $\lambda_1$ being the unique least eigenvalue 0.

Equation~(\ref{4f2}) shows that, in the Rouse model the average displacement depends on only the eigenvalues but not the eigenvectors of matrix $\mathbf{L}$. Notice that, in Eq.~(\ref{4f2}), due to $\lambda_1=0$, the motion of the center of mass has separated automatically from the rest. Moreover, from Eq.~(\ref{4f2}), the behavior of the averaged displacement for
extremely short times and for very long times is obvious.
In the limit of very short times and sufficiently large $N$, $\langle Y(t) \rangle \sim F t/ \zeta$; while for very long times, we have
$\langle Y(t) \rangle \sim F t /(N \zeta)$. The physical explanation is as follows: for very short times only one bead is moving, whereas for very long times the whole GGS diffuses. The above two behaviors are general features for all systems, for a given GGS, its particular topology comes into play only in the intermediate time domain.

In addition to $\langle Y(t) \rangle$, another interesting quantity is the mechanical relaxation form, namely the complex dynamic modulus $G*(\omega)$, or equivalently, its real $G'(\omega)$ and imaginary $G''(\omega)$ components, which are known as the storage and the loss moduli~\cite{Fe80,Wa85}. For very dilute solutions and for $\omega >0$, $G'(\omega)$ and  $G''(\omega)$ for the Rouse model are given by
\begin{equation}\label{4f3}
G'(\omega ) = \frac{\nu k_B T}{N}\sum_{i=2}^N {\frac{{{{(\omega /2\sigma {\lambda _i})}^2}}}{{1 + {{(\omega /2\sigma {\lambda _i})}^2}}}}
   \end{equation}
and
\begin{equation}\label{4f4}
G''(\omega ) = \frac{\nu k_BT}{N}\sum\limits_{i = 2}^N {\frac{{\omega /2\sigma {\lambda _i}}}{{1 + {{(\omega /2\sigma {\lambda _i})}^2}}}}\,,
\end{equation}
where $\nu$ denotes the number of polymer segments (beads) per unit volume.

The relaxation patterns of various polymer systems have been studied in previous works~\cite{GuBl05}, including star polymers~\cite{BiKaBl00,KaBiBl00}, dendrimers~\cite{CaCh97,ChCa99,GaFeRa01,BiKaBl01,GuGoBl02}, hyperbranched polymers~\cite{JuFrFeBl03,BlJuKoFe03,BlFeJuKo04,VoGaJu10}, dual Sierpinski fractals~\cite{BlJu02,JuFrBl02,JuVoBe11}, small-world networks~\cite{JeSoBl00,GuBl01}, and scale-free networks~\cite{Ga12}. Below will compute related relaxation quantities for the treelike small-world polymer networks under consideration.

\subsection{Relaxation patterns }

By substituting the full eigenvalues obtained in section~\ref{Section31} into Eqs.~(\ref{4f2}),~(\ref{4f3}), and~(\ref{4f4}), we can compute, respectively, the averaged displacement $\langle Y(t) \rangle$, the storage modulus $G'(\omega)$ and the loss modulus  $G''(\omega)$ for the relaxation dynamics of the small-world polymer networks $U_g$.

\begin{figure}[h]
\centering
\includegraphics[width=1\linewidth,trim=0 0 0 0]{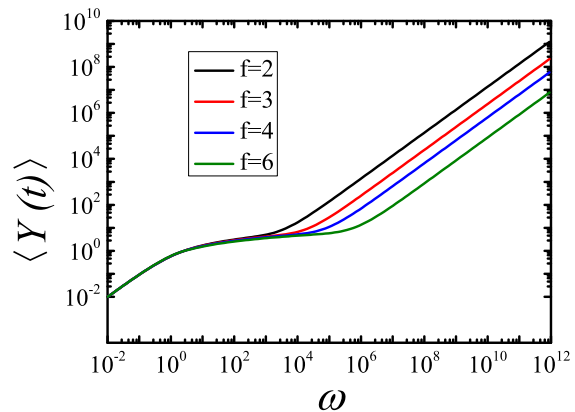}
\caption{(Color online) Averaged monomer displacement $\left\langle {Y(t)} \right\rangle$ for $U_6$ with $f=2,3,4,6$.}\label{AverDispl}
\end{figure}

We begin by focusing on the averaged monomer displacement, $\langle Y(t) \rangle$, given by Eq.~(\ref{4f2}) in which we set $\sigma=1$ and $F/\zeta=1$. In Fig.~\ref{AverDispl} we present in a double logarithmical scale the results of $\langle Y(t) \rangle$ for networks $U_6$ with $f$ ranging from $2$ to $6$. As mentioned above, from Fig.~\ref{AverDispl}, the behavior of $\langle Y(t) \rangle$ for very short and long times are clearly evident, obeying $\langle Y(t) \rangle \sim Ft/\zeta$ and $\langle Y(t) \rangle \sim Ft/(N \zeta)$, respectively. In the region of very short times, only one monomer moves, hence the curves are not dependent on $N$. In contrast, in the domain of very large times, the whole structure drifts, thus the curves depend on $N$: the higher the value of $N$, the slower the limiting long time behavior will be. Typical for the small-world treelike structure is intermediate
time regime, where $\langle Y(t) \rangle$ scales as a power-law behavior with the exponent $\alpha=0.2$ for all $f$, a phenomenon different from that of Vicsek fractals, the exponent of which is related to their spectral dimensions $\tilde{d}=2\ln(f+1)/\ln(3f+3)$.

\begin{figure}[h]
\centering
\includegraphics[width=1\linewidth,trim=0 0 0 0]{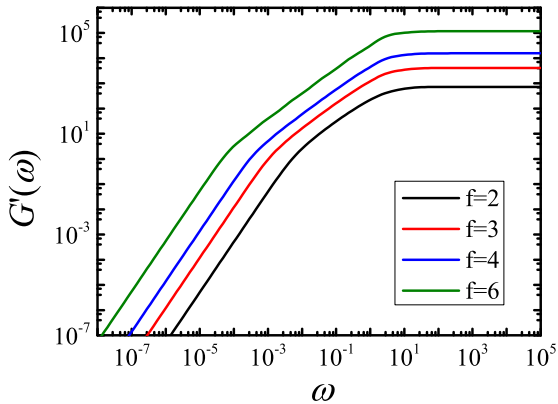}
\caption{(Color online) Storage modulus $G'(\omega)$ for $U_6$ with various $f$.}\label{Storage}
\end{figure}

For the storage modulus $G'(\omega)$, we report the results in Fig.~\ref{Storage}, which is plotted in dimensionless units by setting $\sigma=1$ and $\frac{\nu k_BT}{N}=1$. Figure~\ref{Storage} indicates that in the very low and high frequency limit the storage modulus $G'(\omega)$ exhibit a power-law $\omega^2$ and a plateau, respectively. Both phenomena are the same as those of many different systems. In the intermediate regime the structure being studied play an important role.
For the four cases of $f=2,3,4,6$, we can observe an obvious power-law behavior with an exponent $\alpha'=1$ for all $f$, but the behavior becomes more prominent with $f$ increasing from 2 to 6. It is worth stressing that this result is also different from that for Vicsek fractals~\cite{JuFrFeBl03,BlJuKoFe03,BlFeJuKo04,VoGaJu10}.

\begin{figure}[h]
\centering
\includegraphics[width=1\linewidth,trim=0 0 0 0]{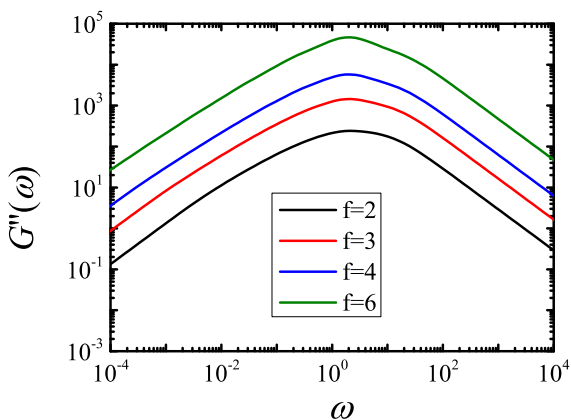}
\caption{(Color online) Loss modulus $G''(\omega)$ for $U_6$ with $f=2,3,4,6$.}\label{Loss}
\end{figure}

For the loss modulus $G''(\omega)$, we plot in a double scale the results in Fig.~\ref{Loss}. As in the case of $G'(\omega)$, we consider  $\sigma=1$ and $\frac{\nu k_BT}{N}=1$. From Fig.~\ref{Loss}, it is easy to notice that for very low frequencies $\omega$, $G''(\omega ) \sim \omega ^1$; and that for very high frequencies $\omega$, $G''(\omega)$ behaves as $G''(\omega) \sim \omega^{-1}$. In the intermediate region,
no power-law behavior is observed, which is in marked contrast to that
corresponding to Vicsek fractals~\cite{BlJuKoFe03,BlFeJuKo04,VoGaJu10}.
It is also important to notice that in the intermediate region,  $G'(\omega)$ and $G''(\omega)$ display different behavior for the small-world structure.

The distinct behaviors for the three quaternities related to relaxation patterns in Viscek fractals and the small-world treelike polymer networks lie in the differences between the two structures. As the name suggests, Viscek fractals are fractals, their relaxation patterns are determine by the fractal dimension and spectral dimension~\cite{JuFrFeBl03,BlJuKoFe03,BlFeJuKo04,VoGaJu10}. For the small-world treelike polymer networks, they are non-fractal, and thus exhibit different relaxation patterns.

\section{Fluorescence depolarization}\label{Section5}

We are now in position to study the dynamics of F\"{o}rster energy transfer over a system of chromophores~\cite{BlVoJuKo05JOL,BlVoJuKo05,GaBl07} positioned at nodes (beads) of the small-world polymer networks. We suppose that the energy can be exchanged only between the nearest neighbors. 
Then, the energy transfer among chromophores located at the nodes of $U_g$ can be described by the following equation
\begin{equation}\label{5f1}
\frac{dP_i(t)}{dt}= \sum^{N_g}_{\substack{j = 1\\ j \ne i}} T_{ij}P_j(t)-\left(\sum^{N_g}_{\substack{j = 1\\ j \ne i}} T_{ij}\right)P_i(t)\,,
\end{equation}
where $P_i(t)$ denotes the probability that node $i$ is excited at time $t$ and $T_{ij}$ represents the transfer rate from node $j$ to node $i$.

As usual, we here separate the radiative delay (equal for all chromophores) from the transfer problem. In fact, the radiative delay only leads to the multiplication of all the $P_i(t)$ by $\exp(-g/\tau_R)$, where $1/\tau_R$ is the radiative decay rate. We presume that all microscopic rates are equal to each other, say $\tilde{k}$, then Eq.~\ref{5f1} becomes
\begin{equation}\label{5f2}
\frac{dP_i(t)}{dt} =-\tilde{k}\,\sum^{N_g}_{\substack{j = 1\\ j \ne i}} L_{ij}P_j(t)- \left(\tilde{k} L_{ii}\right)P_i(t)\,,
\end{equation}
where $L_{ij}$ is the $ij$th entry of Laplacian matrix $\mathbf{L}_g$.

As shown before~\cite{BlVoJuKo05JOL,BlVoJuKo05,GaBl07}, the probability of finding the excitation at time $t$ on the originally excited chromophore, averaged over all possible starting points on $U_g$, is given by
\begin{equation}\label{5f3}
\langle P(t) \rangle  = \frac{1}{N_g}\sum_{i = 1}^{N_g} P_i(t)=\frac{1}{N_g}\sum_{i = 1}^{N_g} \exp (-\tilde{k}\, \lambda_j^{g}\,t)\,,
\end{equation}
which is dependent on all eigenvalues of the Laplacian matrix for $U_g$.

\begin{figure}[h]
\centering
\includegraphics[width=1\linewidth,trim=0 0 0 0]{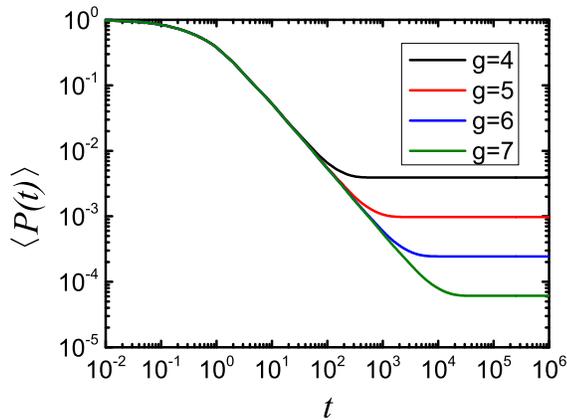}
\caption{(Color online) The average probability $\langle P(t) \rangle$ for $f=3$ and $g=$4, 5, 6, and 7 from above, shown in a log-log scale. }\label{Energy}
\end{figure}

Making use of the eigenvalues obtained in Section~\ref{Section31}, we can evaluate $\langle P(t) \rangle$ for very large networks, without diagonalizing the Laplacian matrix. By setting $\tilde{k}=1$, i.e., by measuring the time in units of $1/\tilde{k}$, we can compute the average probability $\langle P(t) \rangle$ that an initially excited chromophore is excited at time $t$. In Fig.~\ref{Energy}, we present the results for the case $f=3$, with $g$ varying from $g=4$ to $g=7$.

From Fig.~\ref{Energy}, we can see that at very short and very long times, the overall behavior for different $g$ is similar. For example, at long times (depending on the network size), each curve becomes flat, which (in the absence of any radiative decay) is due to the equal distribution of the energy over all nodes in the networks, with each node having a probability of $1/N_g$ of
being excited. We note that similar phenomenon is also observed for Vicsek fractals~\cite{BlVoJuKo05JOL,BlVoJuKo05}. However, at intermediate times, the curves for different $g$ behave quite different, but no scaling is observed, meaning that no curves follow a linear behavior. This phenomenon is as opposed to that for Vicsek fratals, the corresponding curves of which show an obvious algebraic behavior~\cite{BlVoJuKo05JOL,BlVoJuKo05}. The disparity in $\langle P(t) \rangle$ makes it easy to differentiate between Vicsek fractals and the polymer networks studied here.

\section{Conclusions}\label{Section6}

In this paper, we have introduced a class of deterministically growing treelike polymer networks, and shown that they have an exponential-form degree distribution and the small-world characteristic at the same time. We have fully characterized the Laplacian eigenvalues and their corresponding eigenvectors of the networks, which are determined through recursive relations derived from the specific network construction. Using the eigenvalues, we have further studied three representative dynamics for the polymer networks, such as trapping problem, relaxation dynamics in the framework of the GSS, and energy transfer through fluorescence depolarization. Moreover, we have compared the dynamical behaviors with those for Vicsek fractals, which are fundamentally different from each other. Finally, in addition to the aforementioned dynamics, we expect that the obtained eigenvalues and eigenvectors can be adaptable to other dynamics in the small-world networks, e.g., quantum walks~\cite{AhDaZa93,Ke03,AgBlMu08,AgBlMu10,MuBl11}.

\subsection*{Acknowledgment}

This work was supported by the National Natural Science Foundation
of China under Grant Nos. 61074119 and 11275049.

\end{document}